\documentclass[twocolumn]{aastex631}
\usepackage{booktabs}

\newcommand{\bz}{\ensuremath{\langle B_z \rangle}}
\newcommand{\bmax}{\ensuremath{\langle B_z \rangle}^{\rm max}}
\newcommand{\bmin}{\ensuremath{\langle B_z \rangle}^{\rm min}}
\newcommand{\bs}{\ensuremath{\langle \vert B \vert \rangle}}
\newcommand{\Bdip}{\ensuremath{B_{\rm p}}}

\newcommand{\teff}{\ensuremath{T_{\rm eff}}}

\newcommand{\cz}{\ensuremath{c_z}}

\begin{document}

\title{Discovery of magnetically guided metal accretion onto a polluted white dwarf}

\author[0000-0002-7156-8029]{Stefano Bagnulo}
\affiliation{Armagh Observatory \& Planetarium,
College Hill,
Armagh, BT61 9DG, UK}

\author[0000-0003-1748-602X]{Jay Farihi}
\affiliation{Department of Physics and Astronomy,
University College London,
London, WC1E 6BT, UK}

\author[0000-0001-8218-8542]{John D. Landstreet}
\affiliation{Armagh Observatory \& Planetarium,
College Hill,
Armagh, BT61 9DG, UK}
\affiliation{Department of Physics \& Astronomy, University of Western Ontario,
1151 Richmond St. N,
London, N6A 3K7, Ontario, Canada}

\author[0000-0002-9023-7890]{Colin P. Folsom}
\affiliation{Tartu Observatory,
University of Tartu,
Observatooriumi 1,
T\~{o}ravere, 61602, Estonia}

\begin{abstract}
Dynamically active planetary systems orbit a significant fraction of white dwarf stars. These stars often exhibit surface metals accreted from debris disks, which are detected through infrared excess or transiting structures. However, the full journey of a planetesimal from star-grazing orbit to final dissolution in the host star is poorly understood. Here, we report the discovery that the cool metal polluted star WD\,0816--310 has cannibalized heavy elements from a planetary body similar in size to Vesta, and where accretion and horizontal mixing processes have clearly been controlled by the stellar magnetic field. Our observations unveil periodic and synchronized variations in metal line strength and magnetic field intensity, implying a correlation between the local surface density of metals and the magnetic field structure.  Specifically, the data point to a likely persistent concentration of metals near a magnetic pole.  These findings demonstrate that magnetic fields may play a fundamental role in the final stages of exoplanetary bodies that are recycled into their white dwarf hosts.
\end{abstract}


\section{Introduction} \label{Sect_Introduction}
White dwarf stars exhibit multiple hallmarks of their remnant yet vibrant planetary systems, including circumstellar debris that sometimes transits the host \citep{farihi2016,guidry2021}. As the planetary debris is accreted onto the white dwarf, it results in a temporary phase during which photospheric metals can be detected, often referred to as pollution \citep{zuckerman2003}.  Because of the downward settling of heavy elements in their high-gravity atmospheres, white dwarf spectra normally exhibit only hydrogen or helium, but a significant fraction also show traces of these accreted metals \citep{zuckerman2010,koester2014}.  This phenomenon is now well understood to be the result of accretion of planetary material, where the relative elemental abundances measured in polluted white dwarfs provide a gateway to studies of extrasolar geochemistry and planet formation \citep{klein2021,bonsor2023}.

The observed debris disks are attributed to star-grazing planetesimals that are tidally fragmented, but their evolution is understood only in the broadest terms \citep{malamud2020,brouwers2022}.  For example, the necessary drastic compaction of the semimajor axis, from several astronomical units to periastra within a solar radius, is an active area of research in which collisions are likely to play a central role \citep{veras2015,malamud2021}. In some cases magnetic fields may assist in orbital circularization for some disk constituents \citep{hogg2021,zhang2021}.  The accretion of disk particles onto the star is likely to be driven by Poynting-Robertson drag, and possibly enhanced by solid-gas coupling or the collisional destruction of solids \citep{Raf11a,kenyon2017}.  In this process, magnetic fields may play a fundamental role by re-directing the gaseous accretion stream \citep{Metetal12,Faretal18}.

In the local 20\,pc volume of white dwarfs, at least 15\,per cent are metal-polluted \citep{Holletal18}, while over 20\,per cent host a detectable magnetic field \citep{BagLan21}, with a significant overlap between these two populations \citep{Faretal11,Holletal18,BagLan19b,Kawetal19}. It is well known that in main-sequence stars, the presence of a globally organised magnetic field enables and preserves a non-homogeneous distribution of the chemical elements of the photosphere \citep[e.g.][]{DonLan09}. An analogous, but physically distinct phenomenon has been observed in three magnetic white dwarfs that have both hydrogen and helium lines in their spectra \citep{Achetal92,Wesetal01,Mosetal24}, but, perhaps surprisingly, it has hardly been investigated in metal-polluted stars. 

WD\,0816--310 is a helium atmosphere white dwarf at 19.4\,pc distance, exhibiting a DZ spectral type with strong absorption lines from multiple metal species, most prominently Ca, Mg, and Fe \citep{subasavage2008}. Spectropolarimetric observations have detected a variable magnetic field \citep{BagLan19b}. A spectroscopic study by \citet{Kawetal21} argued that the metal abundances in WD\,0816--310 had changed between two observations taken 10\,yr apart, and that this change was probably a consequence of the surface magnetic field.  Here, we report the results of a spectropolarimetric monitoring of this star, aimed at discovering whether there is a relationship between the structure of the magnetic field and the distribution of the metal pollutants on the surface. 
 
\section{Observations}

\begin{figure*}
\begin{center}
\includegraphics[angle=0,width=18cm,trim={1.2cm 2.2cm 1.0cm 0.2cm},clip]{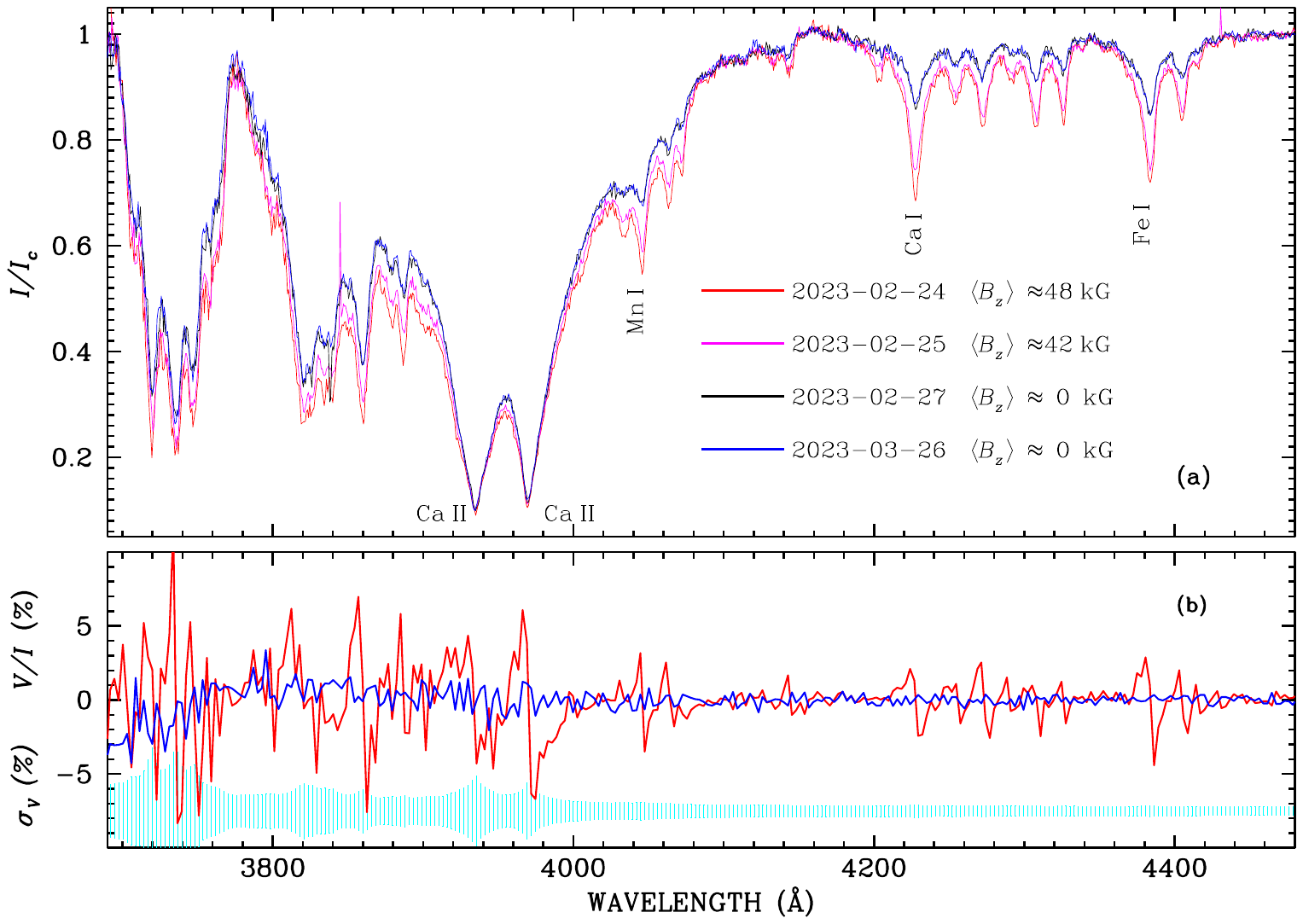}
\end{center}
\caption{\label{Fig_Spectra} Stokes spectra of WD\,0816--310, with flux normalized to the continuum shown in panel (a) and $V/I$ plotted in panel (b). The unsaturated lines are stronger when the longitudinal field is near maximum, and weaker when the field is close to zero.  The presence of a field is revealed by the non-zero Stokes~$V$ profile, as detailed in Appendix~\ref{Sect_Observations}.  For visual clarity, in panel (b) only the $V/I$ profiles obtained on 2023 February 24 and March 26 are shown, and their uncertainties are plotted in cyan, and offset by $-7.5$\%. Line identification from \citet{Kawetal21}.
} 
\end{figure*}

New circular spectropolarimetric observations of WD\,0816--310 were obtained in 2023 February and March with the FORS2 instrument \citep{Appetal98} of the European Southern Observatory Very Large Telescope at Cerro Paranal. FORS2 is a multi-purpose instrument capable of imaging and spectroscopy, equipped with polarimetric optics. Five observations were obtained on four different nights with the grism 1200B, covering the spectral range 3700--5100\,\AA, at resolving power $R\approx 1400$. Data were used to determine the longitudinal component of the magnetic field averaged over the stellar disk, or {\it mean longitudinal field} \bz, and to search for line strength variability (see Appendix~\ref{Sect_Observations}). The dataset shows large changes in the value of both \bz\ and the apparent strength of a number of absorption lines over an interval of a few days (see Figure~\ref{Fig_Spectra}). In particular, the variable lines are wider and deeper when \bz\ is large and positive (2023 February 24 and 25), and weaker when \bz\ is close to zero (2023 February 27 and March 26). The magnetic field measurements, observing dates, and instrument setup are provided in Table~\ref{Table_Log}.
\begin{table*}
\caption{Magnetic field measurements of WD\,0816$-$310 obtained from the (saturated)
Ca {\sc ii} lines and {\it EW} from the lines in the range 4357--4425\,\AA.}\label{Table_Log}
\begin{tabular}{llccr  r@{$\pm$}l r@{$\pm$}l c}
\toprule
INSTR.                           &            
Grism/                           &            
DATE                             &            
UT                               &            
Exp                              &            
\multicolumn{2}{c}{\bz\ (kG)}    &            
\multicolumn{2}{c}{EW (\AA)}     &            
Ref.                            \\            
                                 &            
Grating                          &            
yyyy-mm-dd                       &            
hh:mm                            &            
(s)                              &            
\multicolumn{2}{c}{(Ca\,{\sc ii})}&            
\multicolumn{2}{c}{(Fe\,{\sc i})} &            
                                 \\           
\midrule
FORS1  &  600B  & 2007-10-22 & 08:40 &  960 &$-27.0$& 3.0& 3.35 & 0.1 & BL19      \\ 
ISIS   & R600B  & 2019-04-19 & 21:16 & 3600 &$ 43.8$&15.0& 5.54 & 0.4 & BL19      \\
FORS2  & 1200B  & 2023-02-24 & 05:48 & 2700 &$ 47.9$& 3.9& 5.60 & 0.1 & this work \\
FORS2  & 1200B  & 2023-02-25 & 02:20 & 2700 &$ 42.2$& 3.0& 5.14 & 0.1 & this work \\
FORS2  & 1200B  & 2023-02-25 & 05:12 & 2700 &$ 42.6$& 2.6& 5.04 & 0.1 & this work \\
FORS2  & 1200B  & 2023-02-27 & 05:14 & 2700 &$ -1.4$& 2.6& 3.27 & 0.1 & this work \\
FORS2  & 1200B  & 2023-03-26 & 00:34 & 2700 &$  1.6$& 2.4& 3.17 & 0.1 & this work \\ 
\botrule
\end{tabular}
\tablerefs{BL19: \citet{BagLan19b}.}
\end{table*}

\section{Results}

\subsection{Rotation period and magnetic field morphology}

The measured longitudinal field variability is naturally explained in terms of a static field that is not symmetric about the stellar rotation axis, so that the observer sees a varying field as the star rotates.
The simplest possible magnetic configuration is a dipole field inclined at an angle $\beta$ to the white dwarf rotation axis, which in turn is inclined at an angle $i$ to the line of sight. According to this oblique rotator model, originally introduced to explain the field variations observed in the chemically peculiar stars of the upper main sequence \citep{Stibbs50}, the longitudinal field plotted as a function of rotational phase has a simple sinusoidal shape. In fact, contributions to \bz\ from higher order multipoles tend to cancel out when averaged over the stellar disk, so that \bz\ measurements are mainly sensitive to the contribution from the dipolar component \citep{Bagetal96}, if it exists. 

\begin{figure}
\centering
\includegraphics[width=\columnwidth,trim={0.9cm 9.5cm 0.7cm 3.0cm},clip]{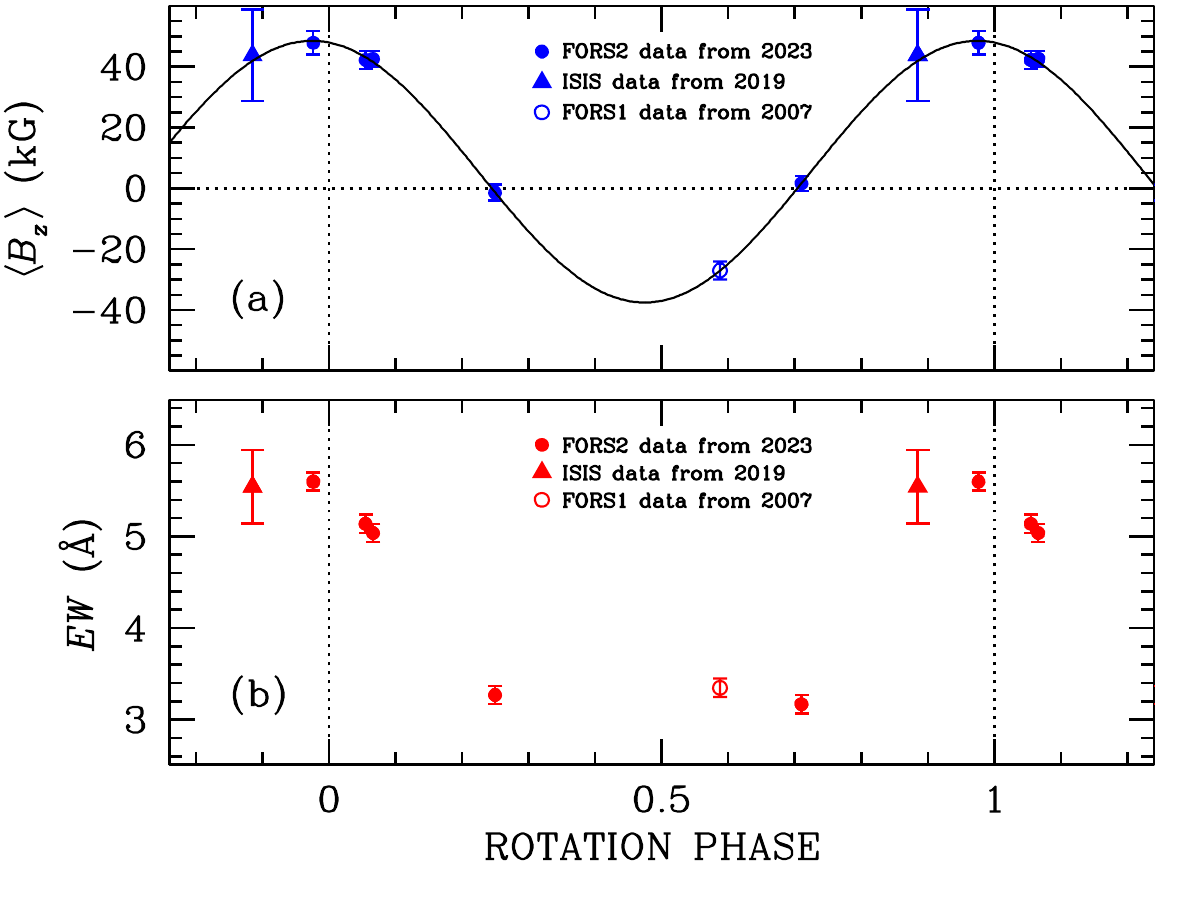}
\caption{Panel (a): the best fit sinusoidal model to the \bz\ data of Table~\ref{Table_Log}, having adopted a rotation period of 10.89426\,d. The zero point of the rotational phase is set arbitrarily at JD=2460000.0. Other values of the rotational period in the range between 2-16\,d lead to equally satisfactory fits with longitudinal curves of similar amplitude. Panel (b): the equivalent width of the lines in the region between 4357 and 4425\,\AA\ phased to the same rotation period and ephemeris as \bz.
}
\label{Fig_WD0816_Bz}
\end{figure}

Sinusoidal curves with several periods between 2 and 16\,d fit the \bz\ values satisfactorily. While the data cannot uniquely determine the spin period, they nevertheless demonstrate that during the rotational cycle, {\it the positive magnetic pole and the magnetic equator cross the line of sight, and at certain rotational phases, the negative magnetic pole transits the visible stellar hemisphere}. In spite of a periodogram with many candidate signals (and aliases of each), a good fit to the varying longitudinal field measurements, assuming a dipolar field, depends only weakly on the choice of the rotational period. Panel (a) of Figure~\ref{Fig_WD0816_Bz} shows the best fit to the \bz\ values obtained by adopting a rotation period of 10.89426\,d. A plausible and simple model consists of a star with $i \approx \beta \approx 70\degr$, and dipolar field strength at the pole $\approx 140$\,kG (see Appendix~\ref{Sect_Orientation}). Panel (b) shows the equivalent width ({\it EW} in Table~\ref{Table_Log}) of the Fe\,{\sc i} lines in the region $4357-4425$\,\AA\ plotted with the same period and ephemeris. Only one spectrum probes the region characterized by a negative field. Additional data are required to find the exact rotation period, to improve the modelling, and to establish how the equivalent widths and \bz\ correlate in the region characterized by a negative longitudinal field.

There are five available sectors of\,{\sc pdcsap} photometry available for WD\,0816--310 from the {\em Transiting Exoplanet Survey Satellite}, (TESS) spanning close to 4\,yr between Sectors 07 and 61.  However, these data reveal no significant periodogram signals above an amplitude of 0.58~per cent, and none that convincingly stand out from adjacent signal peaks. Furthermore, only 10--15~per cent of the\,{\sc pdcsap} flux is from the white dwarf, while over 85~per cent is from other sources.  Thus, no rotation period can be confidently extracted from these data. 

\begin{figure}
\centering
\includegraphics[width=\columnwidth,trim={1.2cm 6.0cm 0.7cm 2.9cm},clip]{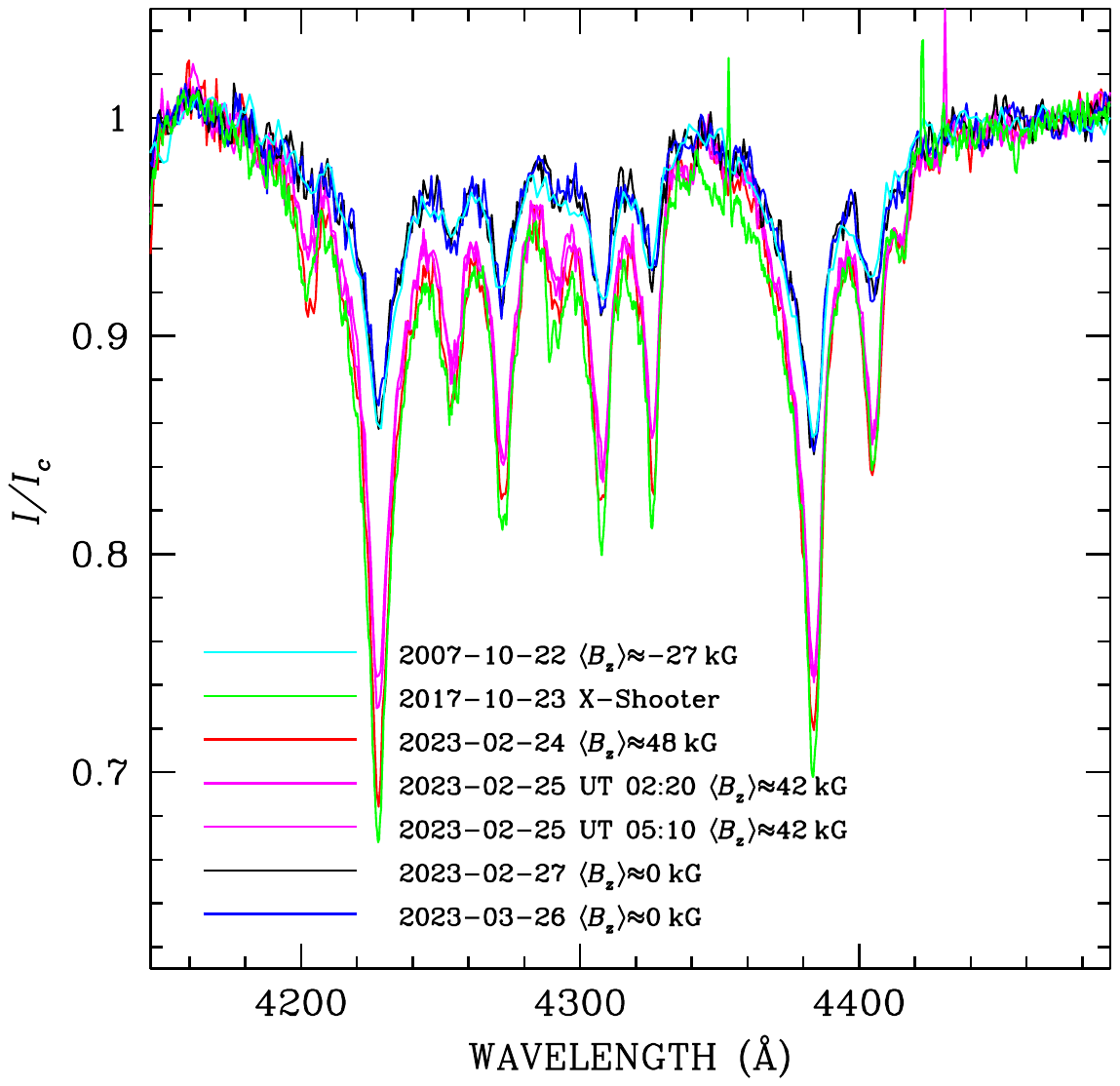}
\caption{All VLT flux spectra of WD\,0816--310 analyzed in this paper, normalized to the continuum, in a wavelength range dominated by Ca\,{\sc i} and Fe\,{\sc i} lines. Note that two spectra obtained on 2023 February 25, nearly overlapping, are plotted using the same color. 
}
\label{Fig_WD0816_X}
\end{figure}

\subsection{Metal line strength variability}

A comparison between a FORS1 spectrum obtained in 2007 and an X-Shooter spectrum obtained in 2017 revealed variations in the strength of the lines corresponding to Na, Mg, Ca, Cr, Mn, Fe, and Ni \citep{Kawetal21}.  Atmospheric modeling demonstrated the elemental abundance changes were clustered around a 0.3\,dex offset between the lower values estimated from the 2007 FORS1 spectrum, and the higher values found with the X-Shooter data taken 10~yr later \citep{Kawetal21}. The essential conclusion drawn from our new data is that these line strength changes occur on a timescale of days, are strongly correlated with the longitudinal component of the magnetic field, and therefore with the rotation of the star. By ruling out the possibility that the abundance changes might be caused by recent accretion (see also Sect.~\ref{Sect_Post_Accretion}), these data represent the first clear detection of abundance spots on a cool polluted magnetic white dwarf.

The FORS2 spectra obtained in 2023 February 24 and 25 show spectral lines of similar strength as those observed with X-Shooter in 2017; the other two recent FORS2 spectra, for which we measure a \bz\ close to zero (2023 February 27 and 2023 March 26), have weaker lines that nearly coincide with those of the earliest FORS1 spectrum (2007), from which we measured a negative \bz\ value. It is clear that the two spectra previously analyzed by \citet{Kawetal21} show roughly the extrema of line strengths for the full available dataset, and therefore define approximately the range of (hemispherically averaged) abundances at the surface of WD\,0816--310 as seen by the observer as the star rotates. All flux spectra in the region 4150--4490\,\AA\ are displayed in Figure~\ref{Fig_WD0816_X}.

Local atmospheric abundance variations (or {\it abundance patches}) are commonly observed in magnetic Ap stars, and one may be tempted to assume that the same mechanism operates in metal polluted white dwarfs, but the physics is entirely distinct. At the surface of an Ap star, abundance patches result from selective radiative levitation of specific ions, guided by the magnetic field, causing large local over-abundances of high-Z elements. In sufficiently cool white dwarfs, radiative forces are no match for gravity, and all heavy elements sink from the atmosphere in a short time.  Metals observed in the atmospheres of white dwarfs are understood to be the result of a relatively recent accretion of material derived from asteroid-like bodies injected onto star-grazing orbits, following the gravitational perturbation by a planet \citep[e.g.][]{bonsor2011}.  The asteroid is then disrupted by tidal forces when it passes within $\sim 1\,R_\odot$ from the star, and the fragments gradually form a circumstellar debris disk whose material is accreted \citep{Jura03}. In $T_{\rm eff}\lesssim11\,000$\,K, non-magnetic white dwarfs, 3D atmospheric models predict that metals will have a homogeneous distribution over the stellar surface \citep{Cunetal21}. Before discussing how the observed abundance inhomogeneities may have been formed and preserved, we examine the accretion history of the star as traced by the multiple heavy elements in the atmosphere.

\subsection{Evidence for a post-accretion epoch}\label{Sect_Post_Accretion}
 
The photospheric metal-to-metal ratios reveal that Mg is notably over-abundant, relative to all heavier elements (Ca, Cr, Mn, Fe, Ni), by a factor of 2--4 as compared to the ratios found in chondritic meteorites, following a trend that correlates with atomic mass and hence diffusion time \citep[Appendix~C;][]{lodders2003}.
This is a well-understood pattern arising from the chemical separation of heavy element species according to their distinct sinking timescales. In the decreasing abundance phase of an accretion episode, infalling material is exhausted, and heavier elements such as Fe sink more rapidly, while the lighter element Mg remains in the photosphere longer \citep{swan2019}. Even under the assumption of steady-state accretion, the Mg overabundance persists, but there is no geochemical pathway to enhance this element over all heavier metals.

\begin{figure}
\begin{center}
\includegraphics[width=\columnwidth,trim={1.2cm 6.0cm 5.4cm 2.9cm},clip]{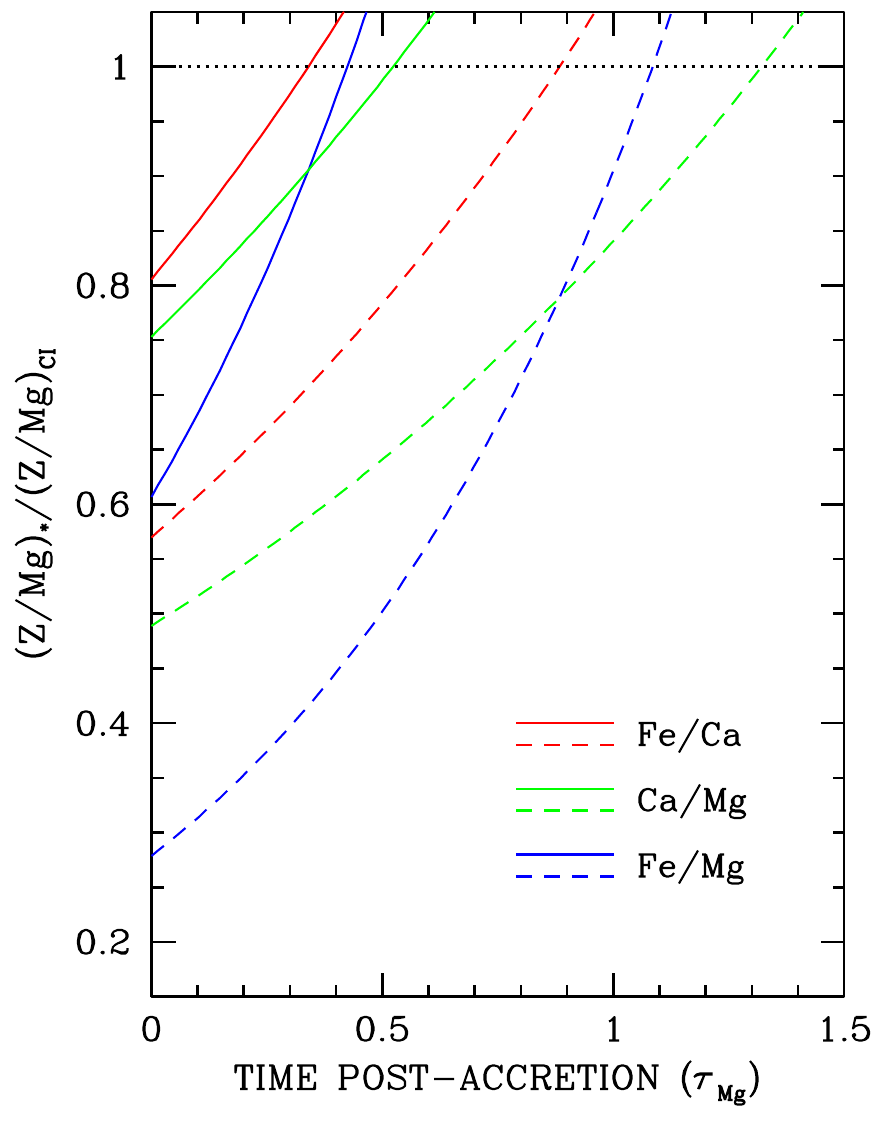}
\vskip 0pt
\caption{Time-reversed, post-accretion evolution of the mass ratios Fe/Ca, Ca/Mg, and Fe/Mg for the accreted source material in WD\,0816--310, normalized to the values of chondritic meteorites \citep{lodders2003}.  Time on the $x-$axis corresponds to $t-t_0$ in Equation~\ref{Eq_Sink}, increases toward earlier epochs, and is normalized by the Mg sinking time.  Two cases are plotted, where the photospheric mass ratios at $t_0$ are the result of 1) an increasing phase, shown by dashed lines, or 2) steady-state accretion, plotted as solid lines.  All three ratios exhibit an under-abundance of the heavier element at the current epoch, but as time moves into the past, these approach and eventually broadly match those of the primitive inner solar system.  Comparing the two plotted models, the solid lines span a narrower range of times when crossing over with the chondritic model, thus supporting a steady state of accretion prior to its cessation.  An additional figure and details are provided in Appendix~C.}
\label{postacc}
\end{center}
\end{figure}

In some polluted white dwarfs, the intrinsic abundances of the debris exhibit departures from chondrites, and in particular can show evidence for differentiated material such as observed in planetary crusts and cores \citep{zuckerman2011,gansicke2012,jura2013}. However, to the limits of the uncertainties, most white dwarfs have accreted planetary bodies that are broadly consistent with chondritic, undifferentiated material \citep{jura2014,xu2019,doyle2023}.  

To model the effects of sinking on the observed heavy elements, we adopt stellar parameters $T_{\rm eff}=6250$\,K and $\log g=8.25$ (cgs), together with the linear average of two sets of metal abundances for Mg, Ca, Cr, Mn, Fe, and Ni calculated by \citet{Kawetal21}. These parameters were then used to define the corresponding atmospheric diffusion models for helium atmosphere white dwarfs with metals (including overshoot and [H/He]~$=-5$; \citealt{koester2020}).

Figure~\ref{postacc} plots the three mass ratios Ca/Mg, Fe/Mg, and Fe/Ca corresponding to the parent body of the accreted material, over time as the three elements sink according to their individual rates.  The plot demonstrates clearly that all three ratios are generally consistent with chondritic values, if accretion ceased between 0.4 and 1.2 Mg sinking timescales ($\tau_{\rm Mg}$) in the past, depending on the prior accretion history.  The best-fitting model corresponds to 0.31~Myr post-accretion, following a steady state. 

Importantly, these three elements behave distinctly in the case of planetary differentiation (e.g.\ by melting): Ca is strongly lithophilic and greatly enhanced in planetary crusts, Fe is strongly siderophilic and dominates cores, while Mg is only mildly lithophilic. These disparate geochemical affinities would readily be revealed in accreted material derived from a differentiated parent body.  As none of these tendencies are observed, the abundance modeling indicates that WD\,0816--310 is polluted by material similar to chondritic meteorites, and that accretion has ceased.

The above analysis is self-consistent and agrees with prior work on polluted white dwarfs in the decreasing abundance phase of accretion \citep[e.g.][]{swan2023}.  On their own, these calculations argue that the magnetic field of WD\,0816--310 has no significant impact on the relative sinking timescales for multiple heavy elements.  This should not be too surprising, as the ratios of sinking timescales for pairs of metals change little over wide ranges of effective temperature and surface gravity \citep{koester2009}, and thus the results of this analysis are robust in the case that magnetism alters the absolute sinking times.  However, even this is unlikely, because atmospheric modeling predicts the field is too weak to significantly impact the size of the convection zone \citep{Bedetal17,Cunetal21}.

The most robust conclusion possible is that the photospheric metals have had at least $0.4~\tau_{\rm Mg}$ or 0.3~Myr to diffuse downward as well as horizontally.  Furthermore, the total mass of metals in the convection zone requires that accretion had been ongoing for a significant period prior to ending.  A minimum mass for the disrupted and accreted parent body can be calculated by assuming the photospheric Fe is 20\,per cent of the total mass as in chondrites, resulting in $2\times10^{21}$\,g residing in the star at present.  If this material has been constantly sinking for roughly one diffusion time, then the minimum mass becomes $5\times10^{21}$\,g.  However, if the star experienced steady-state accretion prior to ceasing, as suggested by the analysis here, then the mass of the parent body can easily be one to two orders of magnitude larger, approaching or exceeding the mass of the asteroid Vesta.  Future improvements in diffusion models will hopefully provide better constraints on accretion history \citep[e.g.][]{heinonen2020}.

\section{The origin of the metal patches}
Spectral variations owing to changes of the local abundance of an element can occur only if the atmosphere is composed of at least two elements. Among white dwarfs, it is quite rare that both hydrogen and helium are comparably abundant and simultaneously exhibit absorption features (e.g.\ spectral type DBA), but in the case of external pollution, there are metals and either hydrogen or helium (or both). Prior to the study of WD\,0816--310, variability of the H/He ratio had been found only in a few DBA white dwarfs.  Among these, Feige\,7, PG\,0853+164 and SDSS J091016.43+210554.2 are the only stars that also show evidence for a (variable) magnetic field \citep{Achetal92,Wesetal01,Mosetal24}. A few other DBA stars, such as G104-27 \citep{Kidetal92}, PG~1210+533 \citep{Beretal94}, WD\,0209$+$085 \citep{Hebetal97}, GD\,323 \citep{Peretal05}, and ZTF\,J203349.8+322901.1 \citep{Caietal23}, show periodic, or at least recurrent, variations in their H/He abundance ratios. In these stars, the presence of a weak magnetic field has often been suggested, but no sensitive magnetic measurements have been made. In addition to these cases, the presence of undetected helium in the atmosphere of the strongly magnetic, cool DAH star G183-35 has been proposed \citep{Kiletal19} to account for the observed variability of the hydrogen line strength, caused by a non-uniform distribution of hydrogen.

Feige~7 and  SDSS J091016.43+210554.2 are  the only white dwarfs for which modelling definitively confirms the direct influence of a magnetic field on localized variations in the atmospheric H/He ratio \citep{Achetal92,Mosetal24}. It is thought that in these hot white dwarfs, the magnetic field completely suppresses the expected surface convection, and thus prevents horizontal mixing, thereby enabling the {\it persistence} of distinct patches with different H/He ratios. However, the mechanism that has led to the formation of the observed patches is not understood.

WD\,0816--310 is the first well-established analogue to Feige\,7, and is the first object found to show abundance patches among cool, metal-rich white dwarfs. In this DZ star, the non-uniformly distributed metals are the remnants of a disrupted asteroid or analogous planetary body, and were deposited from a circumstellar reservoir onto the surface of the white dwarf. We now discuss how metal abundance patches may arise from this specific situation.  

Sufficiently close to the white dwarf, any circumstellar planetary debris will be entirely gaseous, as the equilibrium conditions will eventually exceed the condensation temperatures for highly refractory minerals \citep{rafikov2012,Okuetal23}.  In the absence of a magnetic field, the accreting gas would be expected to arrive on the stellar surface in a band defined by the relative orientation of the disk and the white dwarf rotation axis. The material would then be expected to be spread more or less uniformly over the full white dwarf surface by atmospheric and envelope turbulent convection, on a timescale much shorter than for vertical diffusion \citep[see fig.~11 in][]{Cunetal21}.

The presence of the stellar magnetic field may alter this picture. According to 3D atmospheric models of vertical and horizontal mixing in white dwarfs \citep[][Figure~13]{Cunetal21}, a 100\,kG field is insufficient to inhibit surface convection in a $\teff \approx 6250$\,K, helium atmosphere star such as WD\,0816--310. However, in contrast to these models, the observations show that the $\approx100$\,kG field appears sufficiently strong to inhibit horizontal spreading and preserve the non-uniform arrival pattern of metals, for at least 0.3~Myr (based on Figures~4 and 6, Section~\ref{Sect_Post_Accretion} and the Appendix~\ref{Sect_Ion}).

The mechanism responsible for the initial formation of the abundance patches can be tied to the accretion process as follows.  It is well established that even a weak field will truncate a sufficiently-ionized gas disk and guide the accretion stream to preferentially reach the surface in two patches, one at each magnetic pole \citep{Metetal12,Faretal18}. However, based on the $T_{\rm eff}\approx6250$\,K of WD\,0816--310, it is not obvious that the inward drifting gas will be sufficiently ionized for magnetic control of the accretion flow.  Almost all of the accreting metal atoms have ionization thresholds below 2800\,\AA, where models of cool DZ stars predict flux less than 3\,per cent of that near 4000\,\AA. Therefore, it is unclear if sufficient ultraviolet flux is available to significantly ionize the accretion flow. However, we propose that a powerful ionization multiplier mechanism will be at work in the inner gaseous accretion disk of WD\,0816--310.

Because of near conservation of angular momentum, in the region between the star and disk where solids rapidly sublimate, the circumstellar neutral gas would orbit the white dwarf with a velocity between around 1500 and 3500\,km\,s$^{-1}$.  In other words, metal atoms in orbit carry typical kinetic energies of the order of 1\,MeV.  As soon as one of these metal atoms becomes ionized, it is immediately affected by the magnetic field of the star, which ranges in strength from roughly 1\,kG where the disk transitions from solids to gas, to 100\,kG near the star. As a result of the Lorentz force, the ion will immediately enter a local circular or spiral orbit, with a radius of order meters, around a local field line. It will no longer move with the neutral swarm of atoms, but will at first continue to move at co-rotation speed but in rapidly changing directions. Consequently it will collide with neighbouring neutral streaming particles at high velocity and energy. These collisions will hardly ever restore the rogue ion to neutrality; instead the ion will probably achieve higher ionization states. Furthermore, most of the collisions will result in ionization of the target neutral atoms. The energy available for further ionizations from creation of a single seed ion is enough to ionize thousands of neutral atoms. Many of these new ions will in turn produce still further ions (see Appendix~\ref{Sect_Ion} for additional detail).

In summary, we propose that, even if neither the available ultraviolet flux, nor the gas temperature are sufficient to adequately ionize the disk particles, the presence of a tiny number of atoms ionized by the stellar flux will progressively lead to a substantial level of ionization within the gas disk. Consequently, the accreting stream will be forced to follow the field lines towards the magnetic poles on the surface of the white dwarf. In the polar regions, the inhibition of horizontal mixing will then facilitate the accumulation of metals.

It may be surprising that abundance patches in magnetic, metal-polluted white dwarfs have not been discovered until recently. One reason for this may be the general tendency to obtain only a single spectrum of each target white dwarf, or a lack of detailed comparison between multiple spectra, where available. Furthermore, the presence of abundance patches may well escape detection unless spectra are obtained with high signal-to-noise, e.g.\ S/N $>100$ in the continuum, a level routinely achieved when measuring polarization in spectral lines, but rarely in large spectroscopic surveys.  We predict that by adopting a tailored observing strategy, line variability will become a frequently observed phenomenon in magnetic polluted white dwarfs.

\begin{acknowledgments}
The authors thank Andrew Swan for feedback on the manuscript.  Based on observations obtained with data collected at the Paranal Observatory under program ID 110.243A. JDL acknowledges the financial support of the Natural Sciences and Engineering Research Council of Canada, funding reference number 6377-2016. This research was partially supported by the Munich Institute for Astro-, Particle and BioPhysics (MIAPbP) which is funded by the Deutsche Forschungsgemeinschaft under Germany's Excellence Strategy EXC~2094 – 390783311.
\end{acknowledgments}

\vspace{5mm}
\facilities{VLT:Antu}

\appendix

\section{Observations, data, and magnetic field measurements}\label{Sect_Observations}

The polarization spectra were obtained using the beam swapping technique \citep{Bagetal09}, and reduced using the FORS2 pipeline \citep{Izzetal10}, although the extraction of the wavelength calibrated beams was made with IRAF, and beam recombination was performed using dedicated FORTRAN routines. Intensity (Stokes $I$) spectra were then divided by the spectrum of a featureless DC white dwarf obtained with the identical instrumental setup, and then normalized to a convenient continuum.  Owing to the beam-swapping technique, the measurements of the fraction of circular polarization (Stokes $V$ divided by the intensity $I$) are self-calibrated \citep{Bagetal09}. The sign of Stokes~$V$ is defined as in previous works \citep{BagLan19a}. 

The observed fields are sufficiently weak that the profiles of absorption lines in the ($R \approx 1000$) spectra are virtually unchanged from those expected in the absence of a magnetic field. Thus, the mean longitudinal field \bz\ is determined via a least-squares technique using the correlation between polarization and line slope \citep{Bagetal02} 
\begin{equation}
\frac{V}{I} = -g_\mathrm{eff} \cz \lambda^2 \,\frac{1}{I}\ \frac{\mathrm{d}I}{\mathrm{d}\lambda}\ \bz
\label{Eq_Bz}
\end{equation}
with
\begin{equation}
\cz = \frac{e}{4 \pi m_\mathrm{e} c^2}
\ \ \simeq 4.67 \times 10^{-13}\,\text{\AA}^{-1}\ \mathrm{G}^{-1}
\end{equation}
where $e$ is the electron charge, $m_\mathrm{e}$ the electron mass, $c$ the speed of light, and $g_{\rm eff}$ the effective Land\'e factor.  We note that the derived mean longitudinal field is weighted over the observed stellar hemisphere by the local brightness (which decreases towards the stellar limb), and by the local strength of the spectral line(s) used for the measurement. 

In this work, we analyse the newly obtained FORS2 data, and re-visit observations obtained with FORS2 and the ISIS instrument of the William Herschel Telescope, which were published in a previous discovery paper \citep{BagLan19b}. All of the observing series of sub-exposures were obtained by setting the retarder waveplate at sequential position angles $-45,+45,+45,-45^\circ$. Because white dwarfs may have short rotation periods, we also split each observing series into two pairs of sub-exposures, to search for short-term variability. Because no evidence of variation within a 1\,h observing series was found, the Stokes profiles obtained from each complete series of four sub-exposures were used in this work. 

The method of determining \bz\ through Eq.~\ref{Eq_Bz} has been well tested on many kinds of stars, including white dwarfs \citep{Bagetal12,BagLan18}. Overall, it leads to \bz\ values broadly in agreement with those obtained with other methods, but it is invariably found that results depend on the specific choice of spectral lines or instrument setup \citep{Lanetal14}. In the case of WD\,0816--310, it was found that \bz\ values obtained from the highly saturated Ca\,{\sc ii} lines and \bz\ values obtained from the weaker, variable lines around 4214--4415\,\AA\ show some systematic differences around the epoch of maximum field. For example, for the data obtained on 2023 February 24, the analysis of Ca\,{\sc ii} lines yields $\bz = +48 \pm 4$\,kG, while the analysis of the spectral lines in the range 4214--4415\,\AA\ yields $\bz = +58 \pm 3$\,kG. 

These differences may arise in part from simplifications inherent in the correlation method, which is correct when applied to single lines, but is only approximately correct for a spectral region crowded with numerous blended lines. Furthermore, the Ca\,{\sc ii} lines are strongly saturated, and their shape changes little as the ion abundance varies. While is it not known how the strong saturation of the two Ca\,{\sc ii} lines weight the local contributions to the \bz\ measurement over the visible hemisphere, the similarity of the local Ca\,{\sc ii} line profiles in the observations obtained at different rotation phases may support a roughly uniform weight. The weaker lines around 4214--4415\,\AA\ are instead sensitive 
to small variations of the relevant abundance.  A higher weight will be given to the values of the magnetic fields that characterizes the regions where the elements are more abundant.  As metal elements appear more abundant around the north magnetic poles, they probe with higher weight the regions of the stellar hemisphere characterized by the highest field strength.  

Of course, a significantly better characterization of the magnetic field may be achieved with proper modelling of the observed Stokes parameters, in a way similar to what is currently done for Ap/Bp stars. Such modelling is not possible with the limited set of low-resolution polarization spectra presently available. 

\begin{figure*}
\begin{center}
\includegraphics[angle=0,width=\columnwidth,trim={0.5cm 3.7cm 1.0cm 1.0cm},clip]{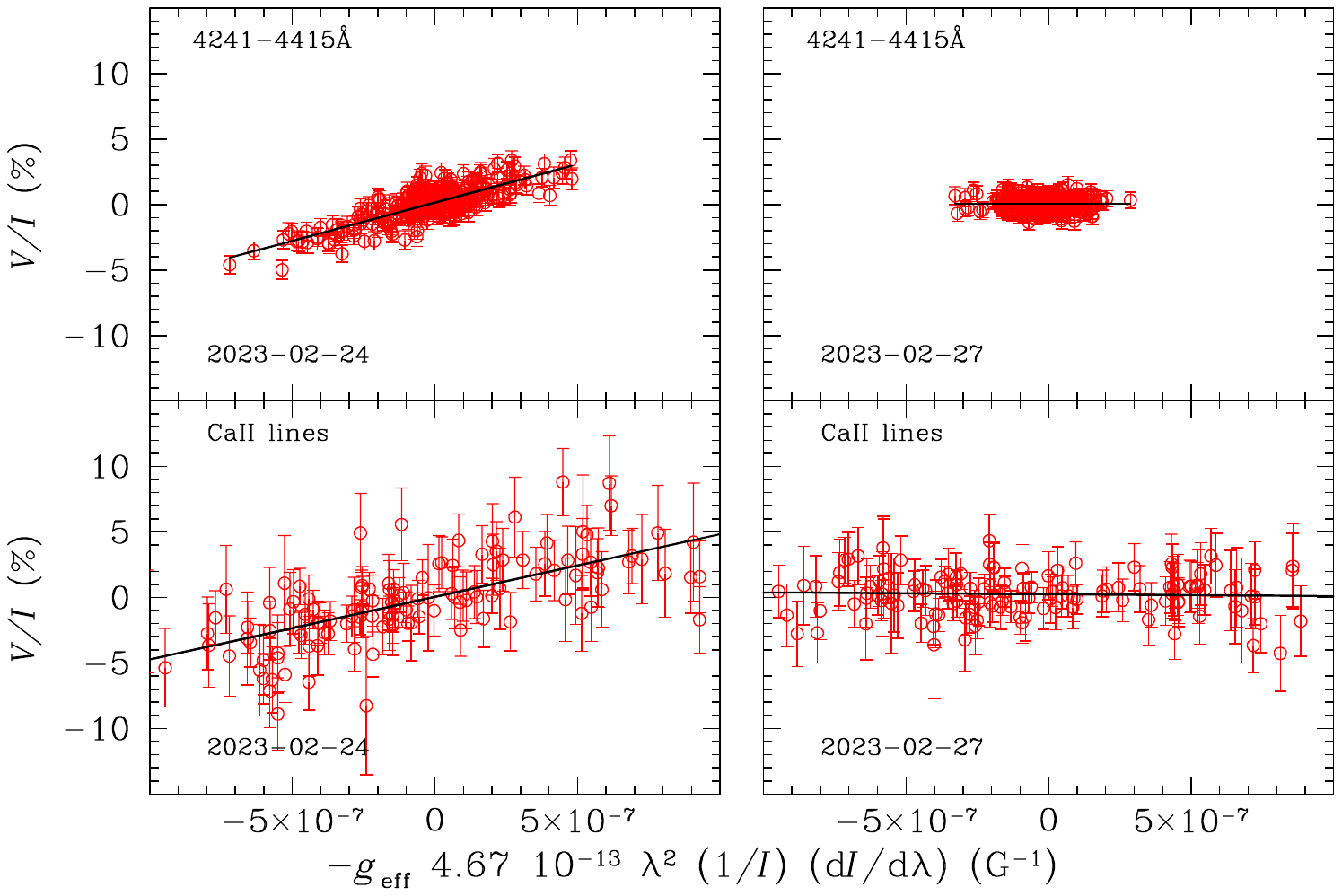}
\end{center}
\caption{\label{Fig_Corr} Eq.~\ref{Eq_Bz} correlations for the observing dates and spectral ranges as given in the figure panel labels. The slopes of the fitted lines are proportional to the mean longitudinal magnetic field.} 
\end{figure*}

In Table~\ref{Table_Log}, we report the field estimates obtained from the combined spectral region 3906-4007\,\AA\ that contains the strong Ca\,{\sc ii} lines, using $g_{\rm eff}=1.25$. Figure~\ref{Fig_Corr} shows the correlation of Eq.~\ref{Eq_Bz} for two observing datasets. These are the values adopted in the paper, keeping in mind that they are field measurements based on a particular spatial sampling as described above. 
 
\section{An approximate model of the stellar magnetic field}
\label{Sect_Orientation}

The simplest possible model of the stellar magnetic field is a dipole of polar field strength \Bdip\ inclined at an obliquity angle $\beta$ to the white dwarf rotation axis, which in turn is inclined at an angle $i$ to the line of sight. Both angles range from 0 
to $180^\circ$ \citep{Lanetal98}, and the variation of the longitudinal field with phase has a simple sinusoidal shape \citep{Stibbs50}
\begin{equation}
\bz \sim 0.31 \Bdip \big[\cos i \cos \beta + \sin i \sin \beta \cos(f-f_0)\big]
\label{Eq_ORM}
\end{equation}
where $f$ is the rotation phase and $f_0$ a zero-phase point such that \bz\ is maximum when $f=f_0$. Equation~\ref{Eq_ORM} shows that if the dipolar field is a reasonable approximation, then the magnetic curve may be fit by a first order Fourier expansion, an assumption that is adopted throughout this work. 

The best fit to the \bz\ data of Table~\ref{Table_Log} is then obtained with a rotation period of 10.89426\,d. This value is not unique, and would be different if \bz\ were obtained from a different interval range. Nevertheless, given the fact that the longitudinal field values has been close to its maximum for about 2 days, and dropped to zero after another 2 days, it is likely that the rotation period is between a few days and $\approx 2$ weeks. 

The \bz\ curve of Eq.~\ref{Eq_ORM} is degenerate, and does not change either if $i$ and $\beta$ values are swapped, nor if $i$ and $\beta$ transform into $180^\circ - i$ and $180^\circ - \beta$, respectively. The latter transformations correspond to a reversal of the rotation direction; therefore, for a physical understanding of the model, it is sufficient to consider only $0^\circ \le i \le 180^\circ$ and $0^\circ \le \beta \le 90^\circ$. A feature of the limited dataset is that $\bmax > 0 > \bmin$ and $\vert \bmax \vert > \vert \bmin \vert $.
This implies that $i + \beta$ must exceed $90^\circ$ in order for the line of sight to intersect both magnetic field hemispheres, and that $i \le 90^\circ$ \citep[see fig.~1c of][]{Lanetal97}.  
We also note that our data allow a rough estimate of the \bz\ curve extrema, which are reached, respectively, at $f-f_0 = 0^\circ$ ($\bz \simeq 48$\,kG), and at $f-f_0 = 180^\circ0$ ($\bz \simeq -39$\,kG). We define
\begin{equation}
r = \frac{\bmin}{\bmax} = \frac{\cos (i + \beta)}{\cos(i - \beta)} 
\label{Eq_r}
\end{equation}
(using Eq.~\ref{Eq_Bz}), and obtain $r \simeq -0.8$. There are an infinite number of pairs ($i$,$\beta$) satisfying this condition. However, some constraints come from the lack of Zeeman splitting in FORS2 data, and from the estimate of the mean field modulus \bs\ obtained from X-Shooter \citep{Kawetal21}, where $\bs \simeq 92$\,kG.  A reasonable approximation of the \bs\ curve for a dipole field is \citep{Henetal77}
\begin{equation}
\bs = 0.64 \Bdip\ \left(1+0.25\ \big[ \cos i  \cos \beta + \sin i \sin \beta \cos (f-f_0)\big]^2\right)\; .
\end{equation}
Therefore, the X-Shooter measured \bs\ value provides the constraint 115\,kG ~$\le \Bdip \le 145$\,kG. Solutions that keep alos the maximum of the longitundinal field $> 40$\,kG are those with $i = 70^\circ \pm 10^\circ$ and $\beta = 70^\circ \mp 10^\circ$, for example $i\simeq \beta \simeq 70^\circ$ and a polar field strength $\Bdip \sim 130-145$\,kG.

A careful comparison of the core shapes of the Ca\,{\sc ii} H and K lines in the FORS 1200B spectra with the X-Shooter spectra indicates that none of our spectra are consistent with $\bs \ge 150$\,kG. This result suggests that by expanding the field structure in a multipole series, the dipole component would still dominate, as the model dipole component accounts for about 100\,kG of the observed H and K line core broadening. 


\begin{figure*}
\begin{center}
\includegraphics[width=14cm,trim={0.5cm 6.0cm 0.0cm 2.9cm},clip]{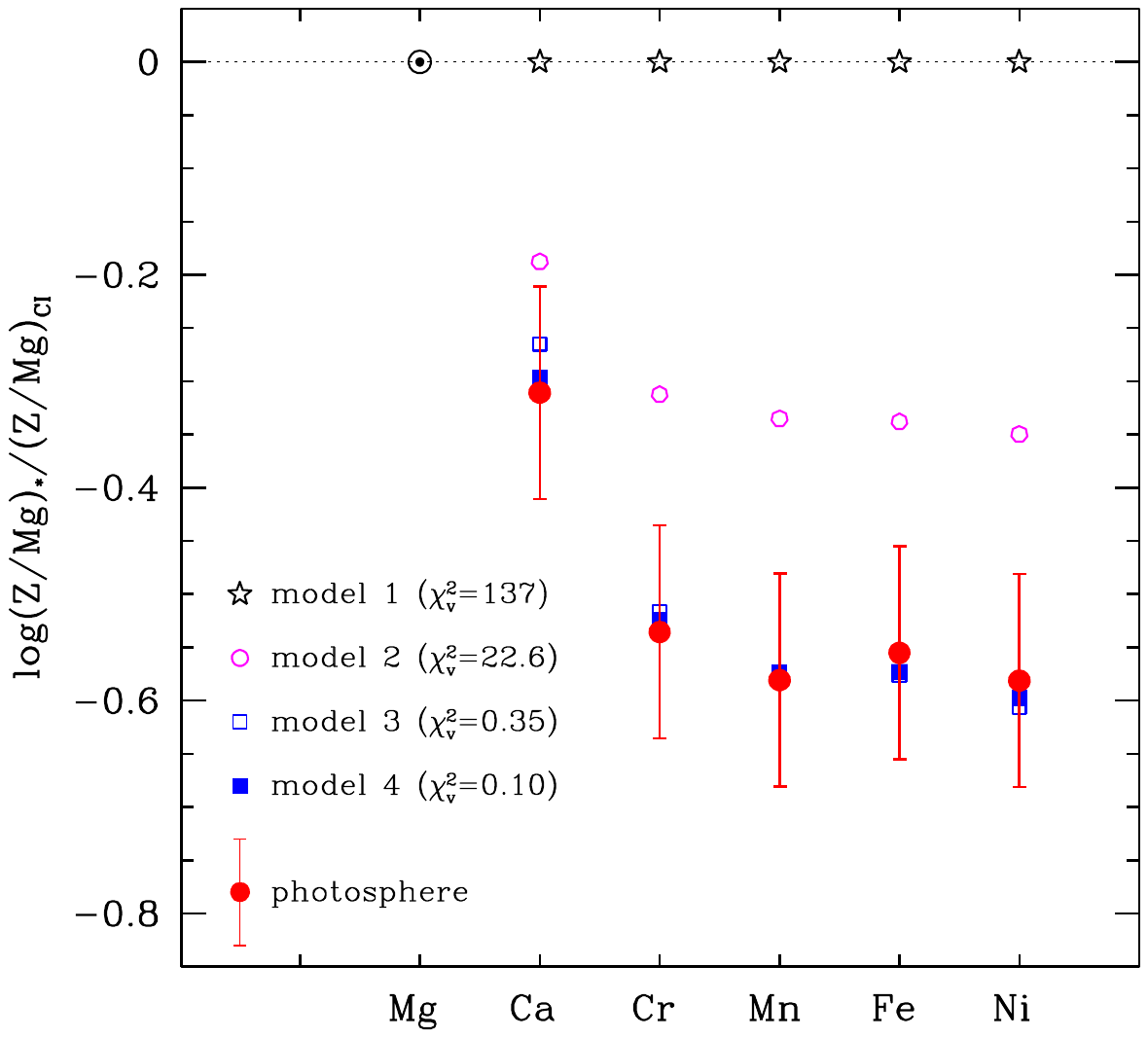}
\end{center}
\caption{\label{Fig_Zmod} The photospheric mass ratios Z/Mg for WD\,0816--310, plotted as filled red circles with error bars, ordered by atomic mass, and compared to models.  The ratios are the linear average of the two sets of elemental abundances determined from atmospheric model fitting \citep[table~3;][]{Kawetal21}, where conservative errors of 0.1\,dex have been adopted based on the internal consistency of the published values. The data are plotted relative to CI chondrites, which are frequently identified as a good match for the parent bodies of white dwarf pollution (see main text).  There are four models shown for comparison, with varying degrees of agreement, all based on the accretion of chondritic material.  Model 1 ()open black stars) is simply chondritic Z/Mg, such as might be directly observed in the star during an increasing abundance phase.  Model 2 (open pink circles) is similar, but where these ratios are modified by steady state accretion.  Model 3 (open blue squares) assumes a decreasing phase of 0.76~Myr follows an increasing phase, and model 4 (filled blue squares) represents a decreasing phase of 0.31~Myr after a steady state.}
\end{figure*}

\section{Accretion-diffusion calculations for heavy elements}

The calculations performed in this paper follow the formal treatment of debris disk accretion and diffusion theory \citep{koester2009,jura2009}.  Specifically, there are three distinct phases that are considered for time $t$ and representative sinking timescale $\tau$; an increasing abundance phase when accretion begins and $t\ll\tau$, followed by a steady-state where accretion is ongoing, and abundances are constant for $t>\tau$, and finally a decreasing abundance phase after accretion has ended (see e.g.\ fig.~2 of \citealt{koester2009}).

There are accurate approximations to the full set of differential equations, and below it is assumed that accretion rate is a constant, and the transition to the decreasing phase is abrupt.  The instantaneous accretion rate for element $z$ in the steady state is given by
\begin{equation}
\dot M_{\rm z} = \frac{X_{\rm z} \ M_{\rm cvz}}{\tau_{\rm z}},
\end{equation}
where
\begin{equation}
X_z = \frac{m_{\rm z}}{m_{\rm He}} \times 10^{\rm [Z/He]}
\end{equation}
is the mass fraction of the element $z$ within $M_{\rm cvz}$, the mass of the convection zone or fully mixed layer of the star, and [Z/He] is its logarithmic number abundance relative to helium.  As detailed in the main paper, the stellar properties ($M_{\rm cvz}, \tau{_z}$) were taken from published model grids, and [Z/He] from published atmospheric model fitting of WD\,0816--310 \citep{Kawetal21}.

In the increasing abundance phase, the mass ratio for any pair of elements in the star, $X_{\rm z1}/X_{\rm z2}$ (e.g.\ Ca/Fe) is identical to that within the parent material, as diffusion has yet to play a role.  Once a steady-state has been established, this ratio is now modified as follows:
\begin{equation}
\frac{\dot M_{\rm z1}}{\dot M_{\rm z2}} = \frac{X_{\rm z1}}{X_{\rm z2}} \frac{\tau_{\rm z2}}{\tau_{\rm z1}},
\label{Eq_SS}
\end{equation}
where the composition of the infalling debris is expressed by the left-hand side.  In the decreasing abundance phase, the mass of each element in the white dwarf gradually diminishes exponentially, so that if accretion ceases at $t_0$, then at some time $t$ later, the photospheric mass ratios are given by
\begin{equation}
\frac{X_{\rm z1}(t-t_0)}{X_{\rm z2}(t-t_0)} = \frac{X_{\rm z1}(t_0)e^{-(t-t_0)/\tau_{\rm z1}}}{X_{\rm z2}(t_0)e^{-(t-t_0)/\tau_{\rm z2}}}.
\label{Eq_Sink}
\end{equation}

Figure~\ref{Fig_Zmod} demonstrates that both the increasing (model 1) and steady-state abundance phases (model 2) are not a good match to the observational data as exemplified by Z/Mg.  Instead, there is a clear and correlated pattern of under-abundance with atomic mass, and thus diffusion time, for both scenarios.  Even for the steady state, the chondritic model yields $\chi^2_\nu=22.5$, and is a poor fit.  These results indicate that WD\,0816--310 is observed in the decreasing abundance phase. 

Adopting a model where chondritic material is allowed to chemically segregate via differential sinking using Equation~\ref{Eq_Sink}, two cases were considered for the initial conditions $X_{\rm z1}(t_0)/X_{\rm z2}(t_0)$: model 3 assumes an increasing phase at $t_0$, whereas model 4 adopts a steady state at $t_0$.  Post-accretion, both scenarios provide excellent fits to the data, where $\chi^2_\nu=0.35$ for model 3 and a decreasing phase lasting 0.76~Myr, and $\chi^2_\nu=0.10$ for model 4 with a decreasing phase of 0.31~Myr.  The latter model is adopted because it provides the closest agreement with the data.


\section{Conditions in accretion region}\label{Sect_Ion}

First, we consider conditions in the accretion region closer to the star than the transition region where solids rapidly sublimate, within about $5\,R_{\star}$, in the absence of a magnetic field.  In this region, we expect a kind of cyclone, with gas flowing around the star in a disk with close to local orbital speed. The circular orbital velocity at distance $r$ from the white dwarf is given by
\begin{equation}
v_{\rm orb} = (G M_{\star}/r)^{1/2}
\end{equation}
With  $M_{\star} = 0.57 M_\odot$ and $R_{\star} = 0.013 R_\odot$, the orbital velocity ranges from about 1300\,km\,s$^{-1}$ in the region where solids rapidly sublimate to about 3000\,km\,s$^{-1}$ close to the star.  With $1.60\times10^{-12}$\,erg\,eV$^{-1}$ and metal atoms of typical atomic weight 35, the kinetic energies of the individual atoms are in the range of about 0.1--2.0\,MeV. 

If no magnetic field is present, one might expect the high-speed gas flow around the star to settle into a disk in approximate hydrostatic equilibrium normal to the plane of the disk. The ion density $n_i(z)$ as a function of distance $z$ from the disk plane, where the ion density is $n_{i0}$, would be then given approximately by 
\begin{equation}
    n_i(z) = n_{i0} {e}^{(-z^2/H^2)}  \; ,
\end{equation}
where $n_{i0}$ is the atom plus ion number density at disk mid-plane; the scale height $H$ is set roughly by the component of the stellar gravity normal to the disk, and is given by 
\begin{equation}
    H^2 = \frac{2kT}{((GM_{\star} A m_{\rm p})/r) } r^2
\end{equation}
where $k$ is the Boltzmann constant, $T$ is the ambient temperature in the gas disk, $G$ is the gravitational constant, $A \approx 35$ is the mean atomic weight of the gas atoms, $m_{\rm p}$ is the proton mass, and $r$ is the distance from the white dwarf. If $d$ is the distance from the centre of the white dwarf in units of the white dwarf radius, we find that $H \sim 3 d^{3/2}$\,km, which ranges from about 3\,km at the stellar surface to 30\,km in the region where solids rapidly sublimate. 

We use the equation of continuity 
\begin{equation}
    \dot{M} \sim n_{i0}(d) (2 \pi R_{\star}) (2 H) v_{\rm acc} (A m_{\rm p})\, ,
\end{equation}
where $v_{\rm acc}$ is the effective radial drift velocity of atoms towards the stellar surface, to estimate the order of magnitude of the ion density $n_{i0}(d)$ in the gaseous accretion radius. Assuming $v_{\rm acc} \sim v_{\rm th}$, where $v_{\rm th}$ is the mean local thermal velocity dispersion of atoms within the accretion disk), we find 
\begin{equation}
    n_{i0}(d) \sim 5\times 10^9 \dot{M_8} d^{-5/2} 
\end{equation}
particles per cubic centimeter, where $\dot{M_8}$ is the mass accretion rate in units of $10^8$\,g\,s$^{-1}$. The particle density in the gaseous disk is thus rather low. 

If the white dwarf has a magnetic field, and the infalling gas completely un-ionized, the situation does not change. The gas circulates in tropical storm mode around the star.  However, if an atom becomes ionized, it immediately feels the Lorentz force and begins to circle a local magnetic field line. The radius $r_{\rm L}$ of the new orbit around the field line is given in Gaussian cgs units by 
\begin{equation}
    r_{\rm L} = A m_{\rm p} v_{\rm orb} c / q B
\end{equation}
where $q$ is the charge of the electron. For a space velocity of $v_{\rm orb} \sim 2000$\,km\,s$^{-1}$, and in a field of 1\,kG, this radius is of the order of 
100\,cm, so the ion is closely confined to a single line of force. 

A newly created ion will thus find itself moving relative to the neutral ions in the gas disk, with a relative velocity of the order of $10^8$\,cm\,s$^{-1}$, effectively carrying a kinetic energy relative to the neutral atoms of the order of 1\,MeV.  If the local neutral particle density is of the order of $10^9$\,cm$^{-3}$, the new ion will suffer a few collisions per second. Most of these collisions will ionize the neutral particles struck, and these new ions will join the set locked to field lines, moving at high velocity relative to the neutrals. It thus appears that formation of even a few ions in the inflowing gas will quickly lead to a cascade of ionization, and the gas will quickly become ionized overall. In this situation, the inflow of gas to the stellar surface is expected to be forced to follow field lines, leading to accretion mainly around the magnetic poles rather than in an annulus around the orbital plane of the accretion disk.

\section*{Data availability}
FORS2 data are available through the \href{http://archive.eso.org}{ESO public archive} (program ID: 110.243A.001, PI=Bagnulo). ISIS data are available at the \href{http://casu.ast.cam.ac.uk/casuadc/}{Astronomical Data Center}.
{\it TESS} data used in this paper can be found in MAST: \dataset[10.17909/3es1-hp33]{http://dx.doi.org/10.17909/3es1-hp33}

\bibliography{sbabib}{}
\bibliographystyle{aasjournal}

\end{document}